# Aggregating Buildings as Virtual Power Plant: Architecture, Implementation Technologies, and Case Studies

Fengji Luo, Ali Dorri, Gianluca Ranzi, Raja Jurdak

*Abstract: The prevalence of distributed generation drives the emergence of the "Virtual Power Plant (VPP)" paradigm. A VPP aggregates distributed energy resources while operating as a conventional power plant. In recent years, several renewable energy sources have been installed in the commercial/industrial/residential building sector and they lend themselves for possible aggregation within the VPP operations. At present, building-side energy resources are usually managed by building energy management systems (BEMSs) with autonomous objectives and policies, and they are usually not setup to be directly controlled for the implementation of a VPP energy management system. This article presents a Building VPP (BVPP) architecture that aggregates building-side energy resources through the communication between the VPP energy management system and autonomous BEMSs. The implementation technologies, key components, and operation modes of BVPPs are discussed. Case studies are reported to demonstrate the effectiveness of BVPP in different operational conditions.*

## Introduction

Modern electrical power grids are experiencing a profound transition from vertical to distributed structure. This is reflected in the increasing number of distributed energy resources (DERs) associated with the power distribution side that include, for example, distributed wind/solar power sources, fossil generation units, plug-in electric vehicles, energy storage systems, various types of flexible and controllable loads. Since the early 21$^{st}$ century, smart grid technology has been developing to accommodate the association of physically dispersed energy resources with the aim of improving the grid operation in an environmental friendly, reliable, efficient, and self-healing manner.

As part of smart grid technology, the concept of "Virtual Power Plant (VPP)" [1] was introduced in the last several years and attracted much attention in both academia and industry. A VPP refers to an entity that aggregates a number of DERs and that possesses the visibility, controllability, and functionality exhibited by conventional power plants. Such aggregation is beneficial to different stakeholders: for DER owners, they can increase the asset value through the markets; the distribution network operator (DSO) and transmission network operator can get increased visibility of DERs for network operations and management; for policy makers, VPP can facilitate the targets for renewable energy deployment and reduction of $CO_2$ emissions; VPP investors and operators can obtain new business opportunities. In the current state-of-the-art, technical research has been extensively conducted on the VPP's classification [1], architecture design [2], and market bidding strategies [3]. In industry, there are two most representative VPP projects. One is the Fenix project in the US [4] that aggregates DERs over a large territory where each DER has its own public coupling point with the public grid. The second project is the AGL VPP project launched by the Australian Renewable Energy Agency in South Australia [5]. Current state-of-the-art VPP prototypes can effectively manage the dispersed energy resources through control channels between the VPP's Energy Management System (EMS) and the DERs.

There is a notable ongoing trend that a large proportion of renewable energy is installed at the building side. The building-side renewable energy thus has large potential to be aggregated in the VPP context. This means that traditionally acting as end energy takers, buildings with renewable energy source installations are now capable of reversely providing large-capacity energy provision support to the grid if they can be aggregately managed properly. With the recent prevalence of Building Energy Management Systems (BEMSs, or known as Home Energy Management Systems (HEMSs) for residential buildings) buildings are equipped with a kind of expert system that takes the role of managing their energy resources based on their own energy management policies and operational requirements. Therefore, due to the penetration of BEMSs/HEMSs, the building-side renewable energy cannot be easily be directly managed by VPP EMS. These technical facts imply that a new VPP protype is required to exploit the aggregation potential of building-side energy resources.

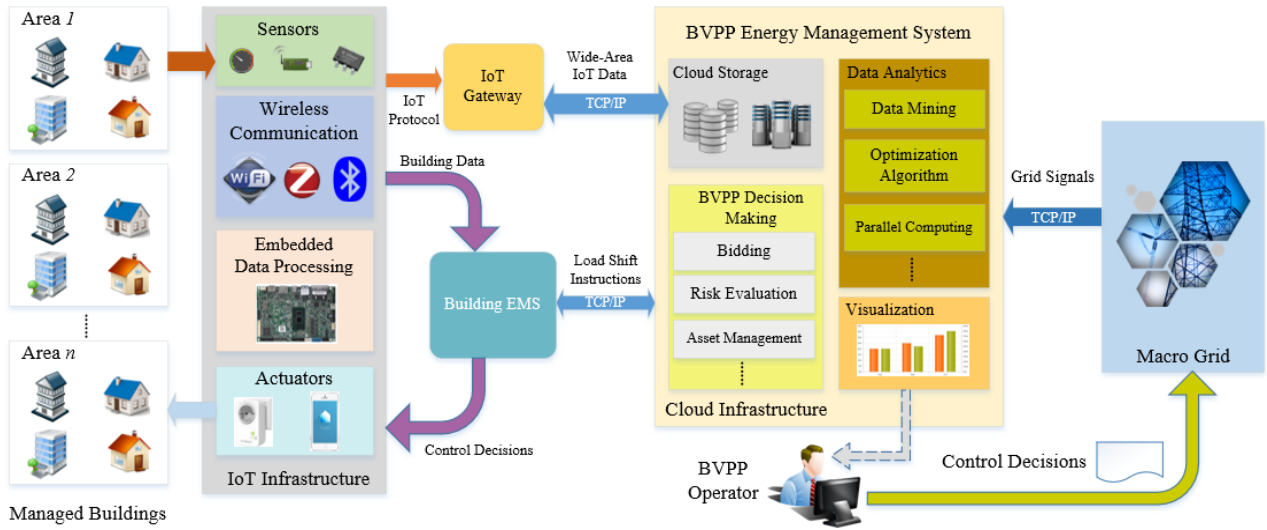

**Figure 1**. Architectural design of BVPP

Motivated by this, this paper proposed a new type of VPP, referring to as the "Building Virtual Power Plant, BVPP". BVPP is based on a decomposed, loosely coupled architecture, in which dispersed building-side energy resources are effectively managed through the interactions and collaborations between BEMSs and the VPP EMS. We firstly present the concept and architectural design of BVPP, then discuss its core implementation technologies and different operation modes. Finally, we design two numerical cases to demonstrate the operation of BVPP in different conditions.

## Architectural Design of Building VPP

The proposed architecture of BVPP is shown in **Figure 1**. Unlike existing VPP prototypes where the VPP EMS directly manages the dispersed energy resources, in BVPP the energy resources in each building are managed by a BEMS. Each BEMS controls on-site energy resources of the building based on the building's autonomous energy management policies and operational requirements. The BEMSs also autonomously interact with external environment (e.g., the grid and utility, GPS system, third-party weather forecasting services, and public platforms).

The BEMSs communicate with the BVPP EMS to achieve certain mutual objectives which cannot be achieved individually. In this way, the energy management of the whole BVPP is decomposed into a coordinated workflow between the BVPP EMS and BEMSs. The obligation of the BVPP EMS is to perform the global control, and send control instructions to the BEMSs. It does not directly control the dispersed energy resources in the building side. Based on the received control instructions from the BVPP EMS, the BEMS controls the on-site building energy resources to satisfy the instruction, while considering the user's requirements and autonomous energy management policies.

The proposed BVPP prototype distinguishes from existing VPP systems for its architectural and technological arrangement. From an architectural viewpoint, the BVPP transforms a centralized VPP energy management framework into a decomposed energy management framework consisting of multiple autonomous, heterogeneous energy entities. From the technological viewpoint, specific energy management application, certain decomposed and distributed algorithms and routines need to be developed to support the proposed architecture. In next two Sections, we will discuss available technologies that can be applied to implement the BVPP architecture, and technically operational modes of BVPP. In Section 5, we will demonstrate a specific energy management application constructed on the proposed BVPP architecture.

## Implementation Technologies of Building VPP

The implementation of BVPP is supported by several key technologies, which are discussed in the following.

### Sensing and Metering Facilities

The operation of BVPP is based on accurately perceiving the status of building energy resources and associated external environments. With an automated architecture established on IP-address based smart meters, current advanced Metering Infrastructure (AMI) can provide BVPP stakeholders and the utility real-time power consumption data of the buildings as well as it can support BEMSs and end users to make energy usage choices based on the incentive signals, prices, and instructions from the utility and BVPP EMS.

At a more fine-grained level, Internet-of-Things (IoT) sensors and networks can deliver more accurate information for BEMSs, BVPP EMS, end users, and other stakeholders to acquire the real-time information of individual devices in buildings as well as external environments. Such ubiquitous information acquisition can provide significant support to improve the BVPP operation.

Another promising technology for monitoring power consumption data at electric appliance-level is the Non-Intrusive Appliance Load Monitoring (NILM) technique [6]. NILM works on the total power consumption data of the building collected by AMI and it infers the most appropriate interpretation on the individual appliance power consumption profile based on the aggregated time series power consumption data. NILM does not rely on customized hardware interfaces and, as a consequence, it is much more economical than installing sub-metering or IoT devices.

*Information Infrastructure*

Different domains of communications are required for the operations of the BVPP. Firstly, machine-to-machine communications exist between BEMS and building energy resources. Several available short-range communication protocols that implement IEEE standards can be applied, such as WiFi (IEEE 802.11 standard), Bluetooth (IEEE 802.15.1 standard), and ZigBee (IEEE 802.15.4 standard). More recent wide area network technologies such as Narrowband IoT (NB-IoT), LoRaWAN, and Sigfox are also candidates for communication within a BVPP.

In BVPP environment, BEMSs would need to communicate and exchange information with other BEMSs. Communication networks covering neighborhood areas with radius of hundreds to thousands of meters are thus required. Candidate communication protocols for this could include WiMax (IEEE 802.16 standard) and Cellular technologies such as 4G/5G.

In a larger scope, BEMSs communicate with the BVPP, and both of them could communicate with the utility, power market, and other party services. Standards TCP/IP protocols can be used for these end-to-end communications.

While the above technologies provide the underlying communication infrastructure, there is still a need for ensuring trust within the BVPP ecosystem, even in the presence of untrusted participants. This is critical to ensure that prediction and control commitments are honored within the network. Immutable, distributed ledge technologies such as Blockchain can provide this trusted environment for BVPPs. Blockchain has been already studied and practically applied in building community environments. Participants can record their interactions through blockchain transactions for an auditable and transparent history. Actionable commitments or predictions can be stored as smart contracts on the blockchain.

*Forecasting Techniques*

BVPPs operate in environments with uncertainties. The stochastic variables in the BVPP environment include renewable power output, occupants' behaviours, air temperature, real-time electricity pricing, and uncontrollable load consumption of buildings. For efficient and correct operational decisions, forecasting techniques need to be implemented and applied.

There have been grain-fined forecasting techniques developed for individual buildings. Unlike traditional regional- or bus-scale forecasting, these techniques take advantages of IoT sensors [7] or NILM [8] to perform forecasting on different objectives in building environments – solar/wind power, occupant lifestyle, building load, etc. These forecasting techniques can provide fundamental support to the decision-making of BVPPs.

*Building Energy Management Systems*

As the BVPP architecture presented earlier, BEMSs are integrated parts of BVPP. BEMS is essentially a kind of expert system that autonomously manages energy resources in buildings. It also acts as a delegation of the user by receiving the user's inputs/requirements and communicate with the utility, BVPP EMS, and other entities to make proper control and scheduling plans for building energy resources.

The design and deployment of BEMS have been widely studied over the years. There are a wide range of BEMSs that are capable of performing different functions. For example, some BEMSs can interface and control the plug-in electric vehicles [9]; some can coordinately manage several kinds of specific controllable appliances [10]. Some other BEMSs can manage energy resources of multiple buildings/houses based on the neighboring communication network [11]. Different kinds of BEMSs can be applied to the implementation of BVPP, based on practical deployment considerations and conditions, including software/hardware interface availability the building energy resources, and users' willingness.

## Operation Modes of Building VPP

Depending on their operational objectives and considerations, BVPPs can operate in different modes as outlined in the following by considering the different operation modes separately.

*Technical BVPP (T-BVPP)*

In the T-BVPP mode, buildings are aggregated as a VPP to provide ancillary services to the external distribution network and wide-area grid. The BVPP EMS communicates with the utility to get awareness on the operational condition of the external grids and it also sends real-time energy generation and consumption information of the buildings to the utility. Based on the condition and requirements of the distribution grid, the BVPP EMS coordinates BEMSs to make control plans on building energy resources and provide ancillary services to the grid including (not limited):

- Peak load reduction - BVPPs can coordinately shift the energy consumption of the managed buildings to

contribute to reduction of regional peak demands of the distribution system, while manning comfort and life convenience of the users within an acceptable range;
- Node voltage/grid frequency regulation - Based on the received real-time information of the grid, BVPPs can reshape the power generation and consumption of buildings to participate in the voltage regulation on different nodes of the distribution network, or contribute to regulate the grid frequency; and
- Self-identification/self-description - BVPPs can provide self-description information of the managed buildings to the utility (e.g. building configuration and energy consumption level, energy efficiency data), and assist the utility to enhance the awareness for the energy entities within the local distribution network.

Controlling energy resources in a single building to provide ancillary services to the local distribution network has been studied extensively. By aggregating buildings as a T-BVPP, the visibility of the dispersed building energy resources can be increased for the system operator, and better coordination among building energy resources can be achieved. In the end, the T-BVPP can increase the controllability of the whole active distribution network.

### Commercial BVPP (C-BVPP)

The renewable energy capacity of a single building would be too small to participate in the energy market due to the capacity barrier. Through the aggregation and management supported by the BVPP EMS, building-side generation sources can gain visibility in the energy market. The BVPP EMS can act as the agent of the buildings to trade surplus renewable energy in the energy market, and all participated buildings can share the profit.

One feature that distinguishes BVPPs from conventional VPPs and generation companies is that BVPPs can not only generate power, but also can provide demand response flexibility through load shifting and reduction performed by BEMSs. This feature reduces the imbalance risk for BVPPs in the energy market, which is mainly incurred by the deviation between the planned and actual power generation.

Overall, running BVPPs as commercial VPPs can maximize the value of building-side energy assets in different energy market environments:

- Participation in the wholesale power market - The BVPP EMS can put offers and bids in the wholesale power market to trade the surplus renewable energy of dispersed buildings as a whole. In this way, the building-side, small-capacity renewable energy sources can be significantly enhanced;
- Participation in the distribution network-side energy market - BVPPs can aggregate energy resources of multiple buildings to sell/purchase energy to/from other buildings through distributed energy markets that emerge in recent year [12]. In this way, the local economics and energy efficiency of the distribution network can be enhanced while the loss of long-distance power transmission; and
- Participation in the regulation market - BVPPs cannot only supply power, but also provide demand response through aggregating the flexible demands in the buildings. These capabilities imply BVPPs have large potential to participate in the regulation market and provide power generation and consumption balancing services to the external grid;

### Occupant-Centric BVPP (O-BVPP)

Buildings are occupant-centric environments. Unlike conventional VPPs, serving end users in the buildings is a critical task of BVPPs. Existing BEMSs can perform energy management and high-quality energy service provision for the users in the managed building based on the considerations, among the others, of energy expense, appliance usage convenience, thermal comfort. Based on this, aggregating buildings as a VPP can further enhance the knowledge and energy sharing among end users, which are hardly achieved by individual BEMSs:

- Local energy sharing among buildings - BVPPs can facilitate different buildings to share surplus renewable energy with each other. Based on certain energy sharing and/or incentive mechanisms, a building can take energy from other buildings that have surplus renewable energy with a price lower than the grid retailing price;
- Enhanced resilience of building communities in abnormal conditions such as grid outage events - In emergent or abnormal conditions, the BVPP EMS can take the role of coordinating the energy resources of the buildings with outages to maximize the social warfare of the BVPP and minimize the resulted disturbance to occupants, e.g. maximize the self-power supply time of the whole building community over the outage period; and
- Energy usage knowledge sharing among end users - By acting as a central communication point, the BVPP EMS can collect the user's energy usage data from individual BEMSs and metering facilities. The BVPP EMS can then extract knowledge from the user data collection and make personalized energy-aware recommendations to the user. For example, by analyzing the user data, the BVPP EMS can recommend a specific user with renewable energy investment plans which are adopted by other users who have similar background and budget with the target user.

Running as O-BVPP is of the key features that distinguishes BVPP from conventional VPPs. Through facilitating energy and knowledge sharing among buildings and end users, the building's energy efficiency and the user's experience is expected to exhibited further improvements.

## Case Studies

In this section, we use two cases to illustrate the operation of BVPP. In the first case, we demonstrate how building occupants can share profits through collaboratively operating the buildings as a commercial BVPP to participate to the energy market. In the second case, we demonstrate how multiple buildings can operate as an occupant-centric BVPP to enable users to share energy usage knowledge with each other and to improve the energy efficiency of the buildings.

The building energy consumption profiles used in this case study are generated using a residential power load generator "SimHouse" (**Figure 2**) developed from the 1st and 3rd authors' lab. The SimHouse is built on an open, scenario-based architecture, and it integrates a set of appliance models and a set of interactive tools to allow users to create different lifestyle models of the occupant. An embedded simulation engine then maps the lifestyle models to physical appliance models to generate power consumption profiles.

### Case 1: Operating C-BVPP in an Energy Market

In this case, we simulate a commercial BVPP by aggregating 50 buildings in the same area to participate in an energy market as independent energy suppliers. The market considered here represents a regional wholesale energy market or a demand-side auction market for energy prosumers. Each building is simulated by 'SimHouse' and is equipped with a certain capacity of rooftop solar panel and has a certain number of controllable appliances (e.g. washing machines, dish washers). A solar power output profile is assigned to each building by multiplying the building's solar panel capacity with a series of solar power output coefficients.

A BESS is assumed to be installed on the bus that hosts the 50 buildings and is owned by the BVPP operator. The BESS is used for absorbing the surplus energy generated by the buildings and discharging the energy back to the grid. The BVPP operator puts bids in the energy market and it is assumed that the BVPP is also a price taker in the market. Based on the forecasted hourly day-ahead market clearing price and the forecasted solar power output of each building, the EMS of the BVPP schedules the energy resources to determine the bid to be put in the market in each hour. The revenue of the BVPP from selling energy in the market is shared among the BVPP operator and all participated buildings according to their contributions to their feed-in energy contribution.

For the purpose of reducing the computational complexity of centrally scheduling the energy resources of a large number of buildings and of ensuring the energy management autonomy of the individual buildings, a decoupled energy management scheme is applied. Firstly, the BVPP EMS delivers the forecasted day-ahead market clearance price to the buildings. Each building's EMS then autonomously schedules the appliances' operations to minimize its energy cost (Eq. (1)) based on the market clearance price and the time-of-use tariff it signs with the electricity retailer.

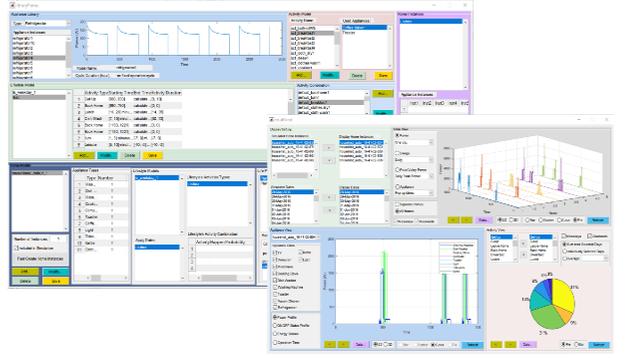

**Figure 2**. Snapshots of the 'SimHouse' simulator.

$$\min F = C_{TOU} - R_{FI} \qquad (1)$$

where $C_{TOU}$ and $R_{FI}$ represent the building's energy cost charged by TOU and the revenue of feeding energy back to the market, respectively. $R_{FI}$ is calculated based on a time-varying feed-in tariff that is set by the BVPP operator that has a smaller value than the market price. Each building then submits its net-load profile to the BVPP EMS. Based on the operation results of the individual buildings, the BVPP EMS optimizes the charging/discharging of the grid battery to determine the bids and to be paid from the market.

Each building schedules its appliances subjected to Eq. (1). **Figure 3** shows the resulted net-load profile of the buildings, in which positive values indicate power consumption and negative values indicate the surplus power production of the buildings. The surplus power of individual buildings is aggregated by the BVPP operator, who optimizes the feeder BESS's charging/discharging decisions to determine the bids submitted to the energy market, which are shown in **Figure 4**. The figure shows that by using the BESS, the BVPP can effectively accommodate the surplus renewable energy generated by the buildings and create more revenues in the market through utilizing the BESS's energy storage capability to optimize the bid in each time interval.

In this study, $340.2 is obtained by selling energy in the market of which 49% corresponds to the BVPP operator's net profit ($165.4) and 51% is shared by the 50 buildings through the time-varying feed-in-tariff scheme ($174.8).

### Case 2: Operating O-BVPP to Facilitate Energy Knowledge Sharing among Occupants

In this case, we demonstrate how a O-BVPP supports knowledge sharing among occupants by providing them with personalized recommendations to improve their energy efficiency. Through load monitoring and IoT sensing technologies, the BVPP EMS can collect the occupants' appliance-level energy usage information. The BVPP EMS then extracts energy efficient collective knowledge from the data and recommends the occupant (called the "target occupant") that has inefficient energy usage habits possible improvements based on the energy-efficient knowledge collected from other occupants (called the "peers") who have similar lifestyles with the target occupant. We use "SimHouse" to generate the major appliances usage profiles of 500 households over 30 days. The major appliances include washing machines, TVs, dish washers, clothes dryers, cooking stoves, and water heaters. These appliances are divided into non-shiftable appliances (lights, TVs, cooking stoves, computers, and water heaters) and shiftable appliances (washing machines, clothes dryers, dish washers, pool pumps, ovens). Non-shiftable appliances represent the occupant's basic lifestyle, while shiftable appliances represent the occupant's Demand Response (DR) potential. The BVPP firstly analyses the 500 households' averaged daily energy usages and costs by using the Fuzzy C-Means (FCM) clustering method to cluster the households into 3 groups (**Figure 5**). Each group is further clustered to identify the households having inefficient energy consumption patterns (red dots in the small figures in **Figure 5**), i.e. the ones which have similar energy consumptions but significantly higher energy costs than the others in the group.

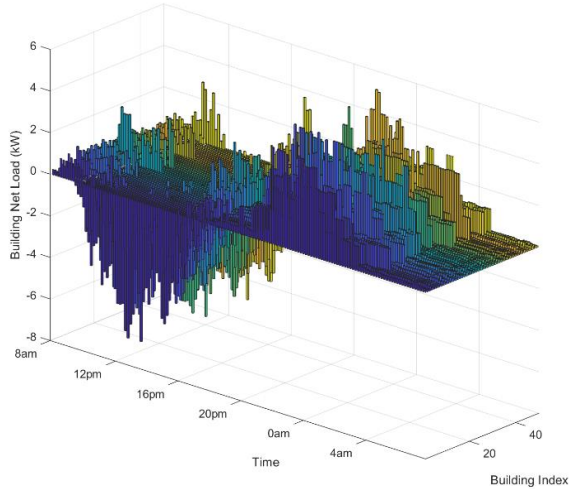

**Figure 3**. Net-load profiles of the buildings.

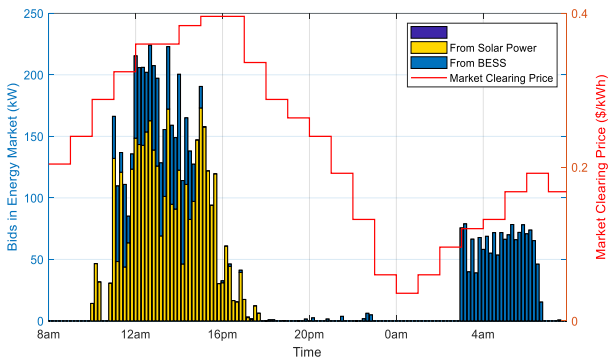

**Figure 4**. Bidding decisions of the BVPP in the energy market.

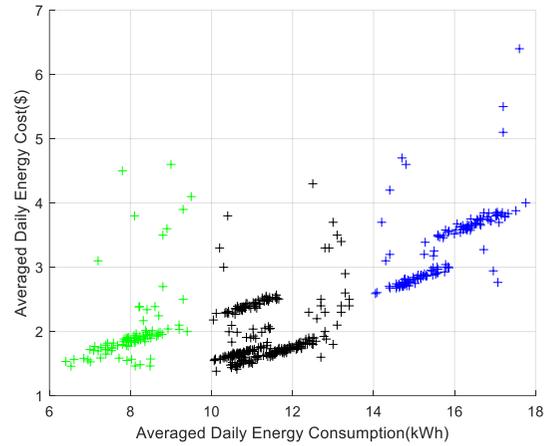

**Figure 5**. Household clustering results.

The lifestyle similarity degree between a target occupant and each household with efficient energy usage habits in the target's cluster is calculated based on the deviation of their non-shiftable appliance usage profiles. Each efficient energy usage plan is then assigned with a 'rating', which is the multiplication of: (a) lifestyle similarity between the target user and the occupants who adopt the efficient energy usage plan; and (b) the cost saving the efficient energy usage plan can achieve compared to the target occupant's current energy usage habit. The top $N$ efficient energy usage plans with largest ratings are then recommended to the target occupant.

A household with inefficient energy usage habits is randomly chosen as the target for demonstration. By applying the above procedures, **Figure 6** shows the top 2 energy usage plans recommended to the target. The figure shows that by analysing the energy usage data of multiple buildings, the BVPP EMS filters out energy usage plans that can well fit the target's lifestyle and are more efficient than the target's energy usage habit. For example, in the second recommended plan, the dish washer, clothes dryer, and washing machine are scheduled to operate in the valley-price period (i.e. midnight) to create more cost saving, while those are scheduled to operate in the evening time in the first plan, so as to avoid noises in the target occupant's sleeping time. Such personalized recommendation is expected to raise the target's awareness on demand response.

By using the same TOU data in Case 1 and applying the recommendation to all 171 occupants with inefficient energy usage habits identified in **Figure 5**, a total amount of $188.1 can be saved for the targets (averaged $1.1 for each one) if they accept the recommended plans.

## Conclusion

This article presents a new type of virtual power plant denoted as 'building virtual power plant (BVPP)' that aims at aggregating building-side energy resources to operate as an integrated power plant. The proposed vision for the BVPP differs from conventional VPPs in following aspects: (1) in-

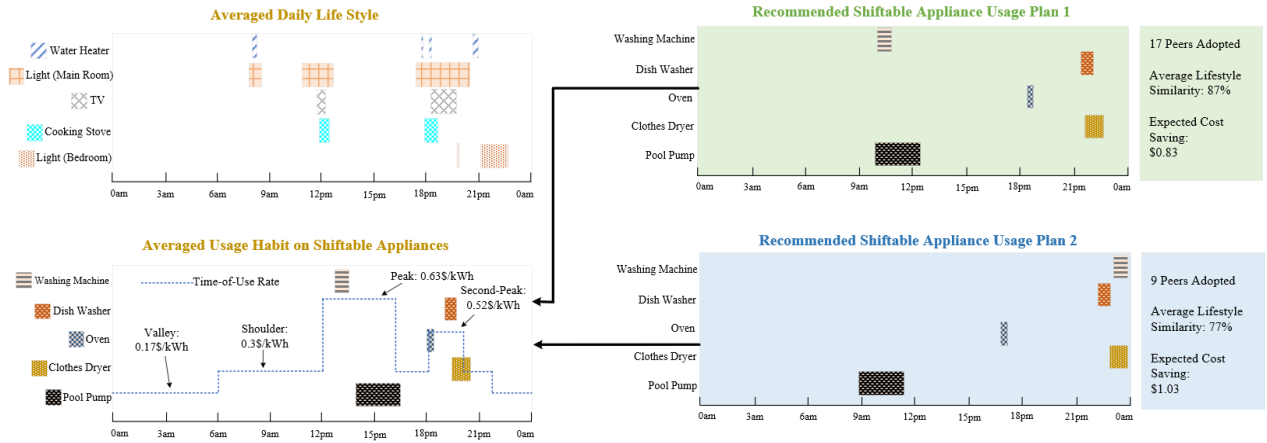

**Figure 5**. Demonstration of energy usage recommendation based on O-BVPP's knowledge aggregation.

-stead of aggregating dispersed generation side resources, the BVPP aggregates energy demand side resources to provide VPP services from the demand side; (2) the BVPP adopts a decomposed architecture in which the BVPP EMS does not directly control building-side energy resources; instead, BEMSs act as middleware and agents to communicate with the BVPP EMS and perform autonomous energy management of individual buildings; and (3) the BVPP operates in a highly occupant-oriented environment.

This article designs two cases to demonstrate the operation of BVPP. In future, more applications are expected to be developed based on BVPP architecture.

## Biographies

Fengji Luo received a Ph.D degree in Electrical Engineering from The University of Newcastle, Australia, 2014. Currently, he is an Academic Fellow in the University of Sydney, Australia. He was a UUKI Rutherford International Researcher at Future Energy Institute, Brunel University London. His research interests include demand side management, smart building, and energy informatics. He has over 130 papers in these areas.

Ali Dorri received his Ph.D degree in Computer Science from the University of New South Wales, Sydney. Currently, he is a research fellow at the Queensland University of Technology. His research interest covers security and privacy concerns in the context of Internet of Things, Wireless Sensor Network, and Vehicular Ad hoc Network.

Gianluca Ranzi was awarded a Ph.D at the University of New South Wales, Australia. He is currently Professor and Director of the Centre for Advanced Structural Engineering at the University of Sydney, Australia. His research interests include high performance building, building-to-grid, and demand side management.

Raja Jurdak is a Professor of Distributed Systems and Chair in Applied Data Sciences at Queensland University of Technology. He received the PhD degree in information and computer science from UC, Irvine. He previously established and led the Distributed Sensing Systems Group at CSIRO. His research interests include trust, mobility and energy-efficiency in networks. Prof. Jurdak has published over 150 peer-reviewed publications. He serves on the editorial board Ad Hoc Networks He is a conjoint professor with the University of New South Wales, and a senior member of the IEEE.